\documentclass[twocolumn,prl,superscriptaddress]{revtex4-2}
\usepackage{amsmath,amssymb,mathrsfs}
\usepackage{natbib}
\usepackage{subfigure}
\usepackage{tabularx}
\usepackage{epsfig}
\usepackage{longtable}
\usepackage{amsfonts}
\usepackage{rotating}
\usepackage{bbold} 
\usepackage{hhline}
\usepackage{braket}
\usepackage{txfonts, comment}

\usepackage[unicode=true,bookmarks=true,bookmarksnumbered=false,bookmarksopen=false,breaklinks=false,pdfborder={0 0 1},backref=false,colorlinks=true]{hyperref}

\hypersetup{linkcolor=magenta,urlcolor=blue,citecolor=blue,pdfstartview={FitH},hyperfootnotes=false,unicode=true}

\def\be{\begin{equation}}
\def\ee{\end{equation}}
\def\bea{\begin{eqnarray}}
\def\eea{\end{eqnarray}}

\def\im{{\rm i}}

\def\Tr{{\rm Tr}}

\newcommand{\norm}[1]{\left\lVert#1\right\rVert}

\begin{document}
\title{The Reservoir Learning Power across Quantum Many-Boby Localization Transition}

\author{Wei Xia}
\affiliation{State Key Laboratory of Surface Physics, Institute of Nanoelectronics and Quantum Computing,
and Department of Physics, Fudan University, Shanghai 200433, China
}

\author{Jie Zou}
\affiliation{State Key Laboratory of Surface Physics, Institute of Nanoelectronics and Quantum Computing,
and Department of Physics, Fudan University, Shanghai 200433, China
}
\author{Xingze Qiu}
\email{xingze@fudan.edu.cn}
\affiliation{State Key Laboratory of Surface Physics, Institute of Nanoelectronics and Quantum Computing,
and Department of Physics, Fudan University, Shanghai 200433, China
}
\affiliation{Shenzhen Institute for Quantum Science and Engineering, Southern University of Science and Technology, Shenzhen 518055, China}

\author{Xiaopeng Li}
\email{xiaopeng\_li@fudan.edu.cn}
\affiliation{State Key Laboratory of Surface Physics, Institute of Nanoelectronics and Quantum Computing,
and Department of Physics, Fudan University, Shanghai 200433, China
}
\affiliation{Shanghai Qi Zhi Institute, AI Tower, Xuhui District, Shanghai 200232, China}

\begin{abstract}
Harnessing the quantum computation power of the present noisy-intermediate-size-quantum devices has received tremendous interest in the last few years. Here we study the learning power of a one-dimensional long-range randomly-coupled quantum spin chain, within the framework of reservoir computing. In time sequence learning tasks, we find the system in the quantum many-body localized (MBL) phase holds long-term memory, which can be attributed to the emergent local integrals of motion. On the other hand, MBL phase does not provide sufficient nonlinearity in learning highly-nonlinear time sequences, which we show in a  parity check task. This is reversed in the quantum ergodic phase, which provides sufficient nonlinearity but compromises memory capacity. In a complex learning task of Mackey-Glass prediction that requires both sufficient memory capacity and nonlinearity, we find optimal learning performance near the MBL-to-ergodic transition. This leads to a guiding principle of quantum reservoir engineering at the edge of quantum ergodicity reaching optimal learning power for generic complex reservoir learning tasks. Our theoretical finding can be readily tested with present experiments. 
\end{abstract}

\date{\today}
\maketitle

{\it Introduction.---}
Recent experimental demonstration of the fascinating quantum computation advantage with superconducting qubits~\cite{2019_Google_Nature} and interfering  photons~\cite{2020_Pan_Science} has triggered monumental research interests in applications of quantum computing. Although a scalable error-corrected quantum computer is still unavailable, a large degree of controllability has been available in a broad spectrum of noise-intermediate-scale quantum (NISQ) devices~\cite{Preskill2018quantumcomputingin,2020_Deutsch_PRXQ,2021_Altman_PRXQuantum},  including atomic~\cite{2017_Bloch_Science}, photonic~\cite{2018_Sciarrino_RPP,2020_Thompson_NatPhys}, 
and solid-state systems~\cite{2020_Oliver_SCqubits,2013_Eriksson_RMP}, whose numerical simulations are exponentially costly in classical computation resources. This suggests compelling quantum computational power embedded the present NISQ devices, of potential importance to applications in science discovery~\cite{2021_Thompson_PRXQ}, cryptography~\cite{2020_Dwave_arXiv}, and optimization~\cite{2020_Hauke_RPP}. How to  harness the quantum computational power and how the computational power manifests in the present NISQ devices demand theoretical investigation. This is particularly crucial for quantum systems lacking full programmability.  

For classical systems, one way to characterize their  computational power is from the learning capability within the framework of reservoir computing~\cite{Nakajima2020}, where the system is interfered by manipulating the input and output only with the system itself untouched as a ``black box". In time sequence learning, it has been found that the complex dynamics in the ``black box" mediates both memory and nonlinearity in mapping the input to the output. It has been demonstrated that a classical dynamical system at the phase edge  of a chaotic region ~\cite{packard1988adaptation,Langton,bertschinger2004rea,LEGENSTEIN2007323,Mushegh_PRX2020} develops optimal learning power as it maintains a balance between integrability and chaoticity that nurture long-term memory and complex nonlinearity.  
 
Generalizing the concept of reservoir computing to quantum dynamical systems leads to quantum reservoir computing (QRC)~\cite{Keisuke2017, Sanjib_npjQI2019, Jiayin2020,  Ghosh2020, nokkala2020gaussian}, 
that does not require precise or full quantum control of the quantum system. This approach provides a plausible route to harness  the ultimate quantum computation power of NISQ devices~\cite{Fujii2020}. However, what quantum systems would exhibit optimal learning power in reservoir computing remains to be understood. 

\begin{figure}[htp]
\vspace{-0.2 cm} 
\includegraphics[width=0.9\linewidth]{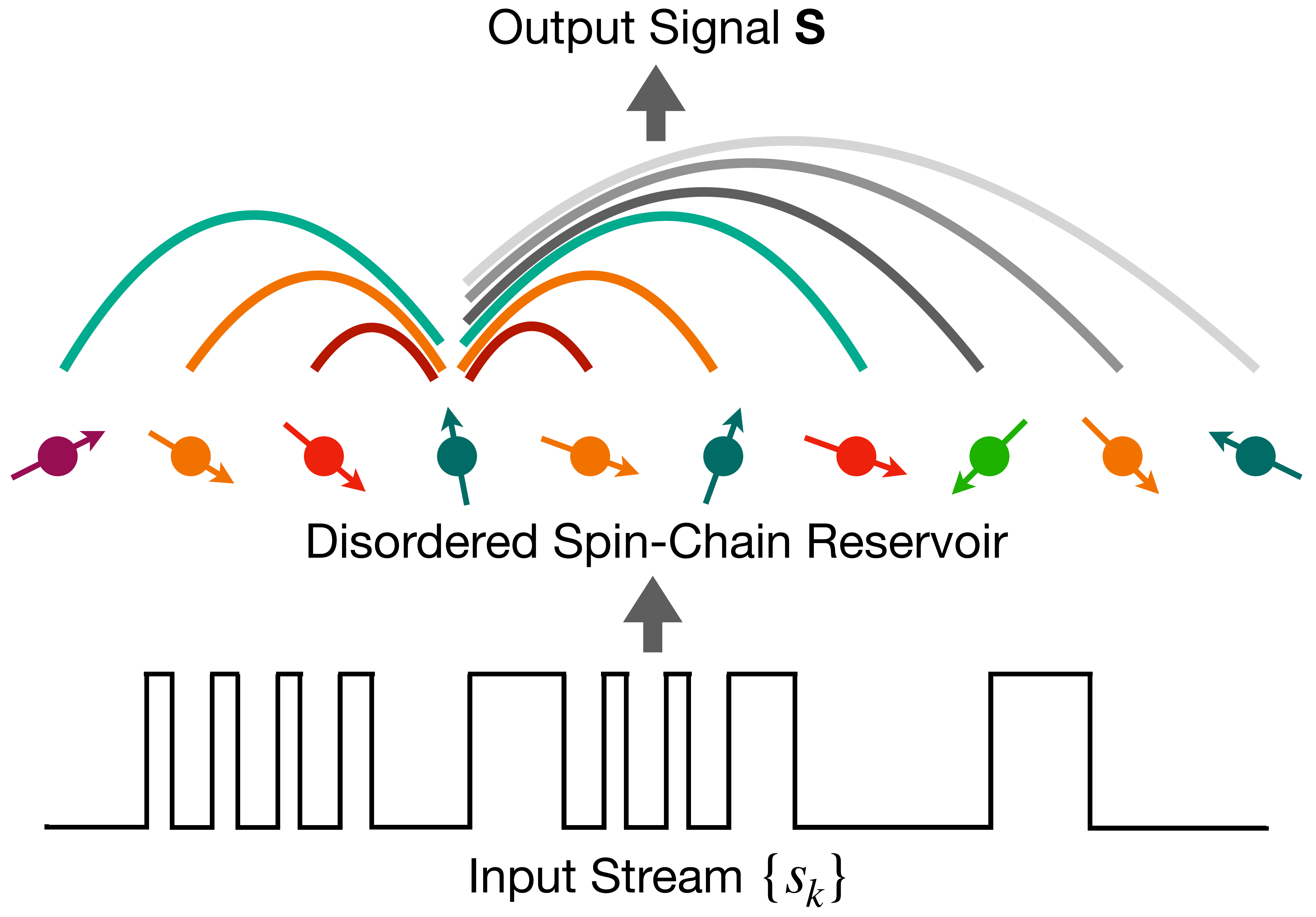}
\caption{Illustration of quantum reservoir computing with a disordered quantum spin chain with long-range interactions. 
The information from the input stream $\{s_k\}$ is fed into the reservoir for information processing. The output signal ${\bf S}$ is collected by quantum measurements on the reservoir qubits. 
}
\label{fig:QRC}
\end{figure}

In this work, we study the learning power of a one-dimensional quantum reservoir composed of long-range randomly-coupled qubits (Fig.~\ref{fig:QRC}).  This system has a quantum many-body localization (MBL) transition with varying the disorder strength. We investigate the learning performance on two computation tasks of Boolean function emulation including the short-term memory (STM), and the parity check (PC) tasks, which characterize the memory capacity and the amount of nonlinearity of the quantum reservoir, respectively~\cite{Keisuke2017}. In the MBL phase, learning the STM task outperforms  the PC task significantly, which we attribute to the extensive number of emergent local integrals of motion in the MBL phase~\cite{2013_Serbyn_PRL,2014_Huse_MBL_PRB,Ros2015420,chandran2015}. In the quantum ergodic phase~\cite{Abanin_RMP2019}, we find the opposite, which implies the quantum dynamics in the ergodic phase mediates larger amount of nonlinearity, more resembling the chaotic dynamics in classical systems. We further examine  the learning power of the disordered quantum many-body system on the Mackey-Glass (MG) prediction task~\cite{mackey1977oscillation}, whose performance requires both memory and nonlinearity. In this task,  we find that the learning power develops a peak around the quantum MBL-to-ergodic  phase transition. This work would shed light on the quantum reservoir engineering to invest most quantum learning power for complex reservoir computing tasks.

{\it The Quantum Reservoir Computing Setup.---}
In QRC~\cite{Keisuke2017}, the system  is initialized as an infinite temperature state at time $t=0$.  
The quantum reservoir dynamics corresponds to a unitary time evolution of $N$ qubits described by a time-dependent density operator,  $\rho(t)$, sequentially interrupted by an input signal  $\{s_k\}^M_{k=1}$, where $s_k$ takes binary numbers $0$ and $1$ for a discrete signal, or values in between in a continuous setting.
The time evolution is chopped into multiple $\tau$-durations, i.e., $t = k\tau$, with each duration further split into $V$ subintervals. At time $t=k\tau$, the time evolution is interrupted by the input as, 
$\rho(t)\mapsto \rho(t+0^+)=\rho_{s_k}^{[1]}\otimes\Tr_1\left[\rho(t)\right]$, 
where $\Tr_1$ is a partial trace with respect to the first qubit, 
and the state of the first qubit is modified to $\rho_{s_k}^{[1]} \to \ket{\psi_{s_k}}\bra{\psi_{s_k}}$, with $\ket{\psi_{s_k}} = \sqrt{1-s_k}\ket{+}+\sqrt{s_k}\ket{-}$, and $\ket{\pm}$, the eigenstates of the Pauli-$X$ operator. Measurements  are performed in the Pauli-$X$ basis at the end of each subinterval. The resulting output signal is denoted as ${\bf S}$, with elements 
$S_{k,v,i}=\left(1+\Tr\left[\,\rho\left(k\tau+\tau v/V\right)X_i\, \right]\right)/2$, $v$ and $i$ indexing subintervals and qubits, respectively.

In learning a time-sequence defined by a nonlinear function, 
$y_{k'} ^\star = f(\{s_{k\le k'}\}; k')$, 
we cast the time steps into three groups $\{K_1, K_2, K_3\}$, 
with the casted time sequence and the reservoir output signal represented by $\{{\bf y}_1^\star, {\bf y}_2^\star, {\bf y}_3^\star \}$, 
and $\{{\bf S}_1, {\bf S}_2, {\bf S}_3\}$ accordingly. The starting time steps $K_1$ are discarded to remove the effect of  initialization. The following time steps $K_2$  are used for training a linear regression model, 
$y_k= \sum_{v,i} S_{k, v, i} W_{v,i} + B_k $ with $W$s and $B$s fitting parameters introduced to minimize $\norm{y-y^{\star}}$. 
The final time steps $K_3$ are for prediction, using the reservoir output ${\bf S}_3$ and the trained linear regression model, by which a time sequence ${\bf y}_3$ is predicted. The performance of the reservoir computing is quantified  by a normalized covariance, 
$C = \rm{cov}\left(\mathbf{y}_3^\star,{\mathbf{y}}_3\right)\Big/\left[\sigma\left(\mathbf{y}_3^\star \right)\sigma\left({\mathbf{y}}_3\right)\right]$, 
with $\sigma$ the standard deviation. Its average $\overline{C}$ is obtained by sampling distinctive  input-signals ($s_k$).

{\it Disordered Quantum Spin Chain Reservoir.---}
We consider a one-dimensional long-ranged coupled transverse-field Ising model as the quantum reservoir (Fig.~\ref{fig:QRC}), whose dynamics is governed by the Hamiltonian, 
\be
\textstyle H = \sum_{1\leqslant i<j\leqslant N}J_{ij}X_{i}X_{j}+\frac{1}{2}\sum^N_{i=1}\left(B+\phi_i \right)Z_{i} \, ,
\label{eq:Ham}
\ee
Here, $X$ and $Z$ are the Pauli operators, $J_{ij}=J_0|i-j|^{-\alpha}$ represent the long-range power-law decaying Ising couplings, and the transverse field contains a background constant part $B$, and a random part, $\phi_i$ drawn from  $\left(-\frac{W}{2},\frac{W}{2}\right)$ according to a uniform distribution. Hereafter, the coupling strength $J_0$ is set as energy unit. In this work, we choose $B = 4$, and the main findings  are largely independent of this parameter choice. 
This disorder spin model has a MBL to quantum ergodic transition, at a critical disorder strength $W_{\rm C}$ \cite{Maksymov_PRB2020}. Long-range coupling is considered here to support sufficiently complex dynamics for QRC. 
To characterize the learning power, we average over the disorder samples and input signals. 

The  QRC model in Eq.~\eqref{eq:Ham} can be realized in experiments with  the present NISQ devices including NMR~\cite{2005_Chuang_RMP}, trapped ions~\cite{2010_Duan_RMP}, and Rydberg atoms~\cite{2020_Browaeys_NatPhys}. With NMR, the required Hamiltonian is realizable by pulse shaping~\cite{2000_Leung_PRA,2017_Du_NMROTOC}. With trapped ions, the model has already been used to investigate the MBL transition~\cite{Smith_NatPhys2016} and discrete time crystals~\cite{2017_Monroe_timecrystal}. With Rydberg atoms, the Ising couplings take specific forms with exponents $\alpha =3, 6$~\cite{2020_Browaeys_NatPhys, wu2020concise, Morgado2020Rydberg}, and generic long-range interacting models can be engineered with the quantum wiring scheme~\cite{2020_Qiu_PRXQ}. In our numerical results to present below, we neglect the backaction effects in the measurements to comply with the NMR system~\cite{2005_Chuang_RMP}. We also confirm the main findings still hold when the  backaction is incorporated.

\begin{figure}[htp]
\includegraphics[width=0.9\linewidth]{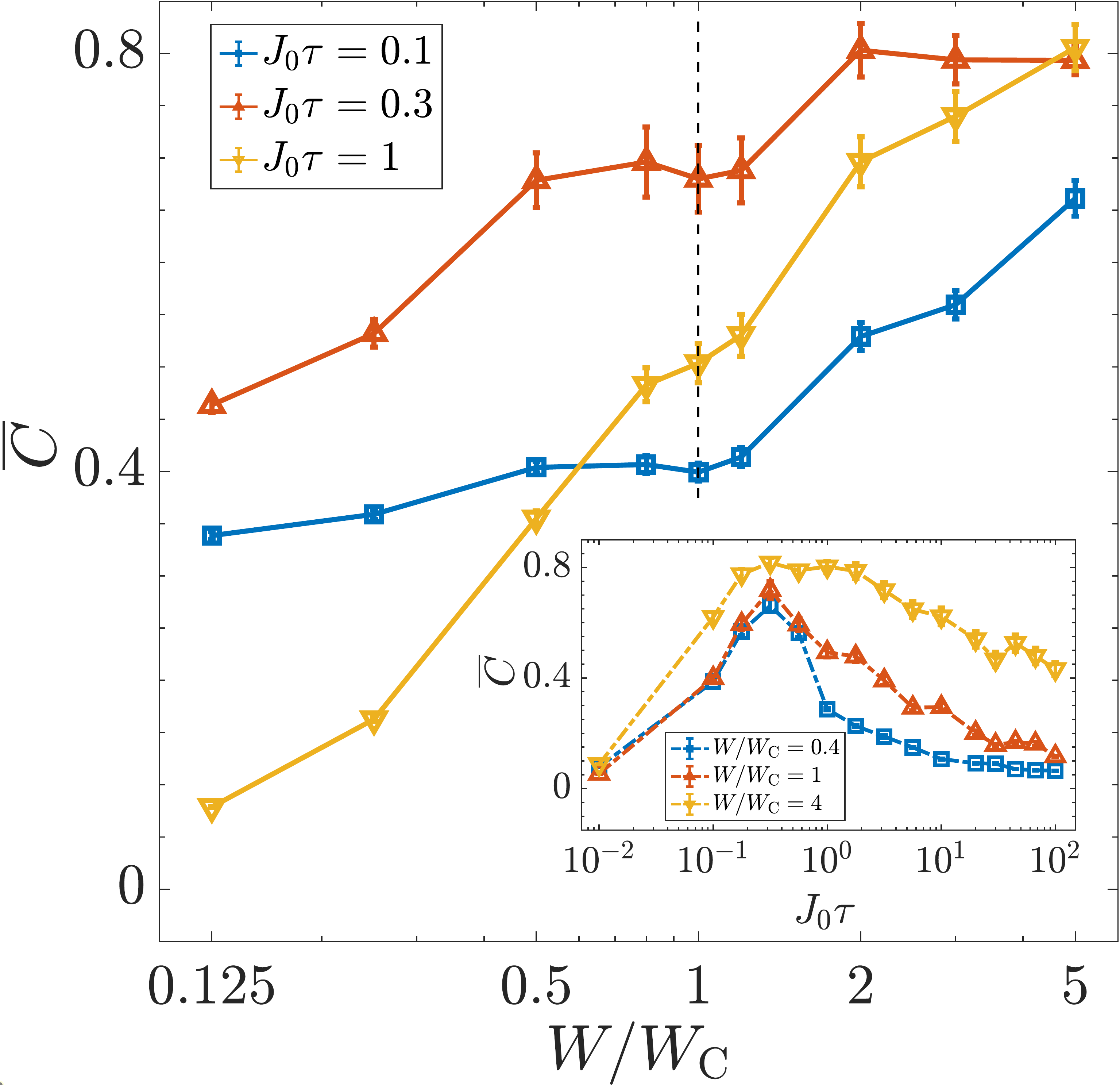}
\caption{The quantum reservoir learning performance on the short-term memory task with varying disorder strength. 
The normalized covariance $\overline{C}$ characterizes the degree of how the reservoir prediction matches the targeting time sequence. The learning performance monotonically improves with increasing disorder strength ($W$). This is found for a broad range of time duration $\tau$ (see main text). 
The inset shows the dependence of $\overline{C}$ on $\tau$. 
The reservoir is in the MBL (ergodic) phase when $W/W_{\rm C}>1$ $\left(W/W_{\rm C}<1\right)$, with the phase boundary marked by the black ``dashed" line. Each task instance consists of 5000 time steps, with the first $1000$ steps discarded, the middle $3000$ steps used for training, and the last $1000$ steps for reservoir prediction and performance evaluation.
Here, we set the qubit number $N=10$, the number of subintervals $V=10$, the time delay $k_\Delta=16$, and the interaction exponent $\alpha=0.4$. The results are averaged over $100$ random samples, with the statistical error provided in the plot.  
} 
\label{fig:STM}
\end{figure}

{\it Short-Term Memory Task.---}
We first evaluate the memory capacity of our quantum spin-chain  reservoir through STM task \cite{Keisuke2017}, where the targeting time-sequence function is 
$
y^\star_k = s_{k-k_\Delta} 
$ 
with $k_{\Delta}$ representing the time delay. The learning performance on this linear task reflects the reservoir memory capacity. 
Fig.~\ref{fig:STM} shows the QRC performance across the MBL transition. We find that the learning performance has a non-monotonic dependence on the time duration $\tau$, reaching an optimum at a certain intermediate value $\tau_c$.
The increase of $\overline{C}$ at small $\tau$ is because the reservoir dynamics is essentially frozen with a too small choice of $\tau$, and that the  information injected through the first qubit is washed away by the subsequent inputs without capability of sharing and storing the information collectively in the reservoir. 
At $\tau > \tau_c$, the decrease of $\overline{C}$ with $\tau$ happens because the information would become more hidden in many-body correlations~\cite{2018_Stanford_JHEP,2019_Li_PRA}, which cannot be extracted by the local measurements.  
An intermediate value of $\tau$ should be used in order to maximize the learning power of the reservoir. 
As we increase the disorder strength, the learning performance exhibits a monotonic increase, demonstrating an apparent advantage of the MBL phase ($W>W_C$) in memory holding, compared with the quantum ergodic phase ($W<W_C$).  We attribute this to the emergence of local integrals of motion in the MBL phase~\cite{2013_Serbyn_PRL,2014_Huse_MBL_PRB,Ros2015420,chandran2015}. 
With these conserved quantities, the information stored in the Pauli-$X$ basis scrambles very slowly through a logarithmic dephasing process in quantum  dynamics~\cite{2012_Moore_PRL,2013_Abanin_PRL,2014_Demler_PRL}.

\begin{figure}[htp]
\includegraphics[width=0.9\linewidth]{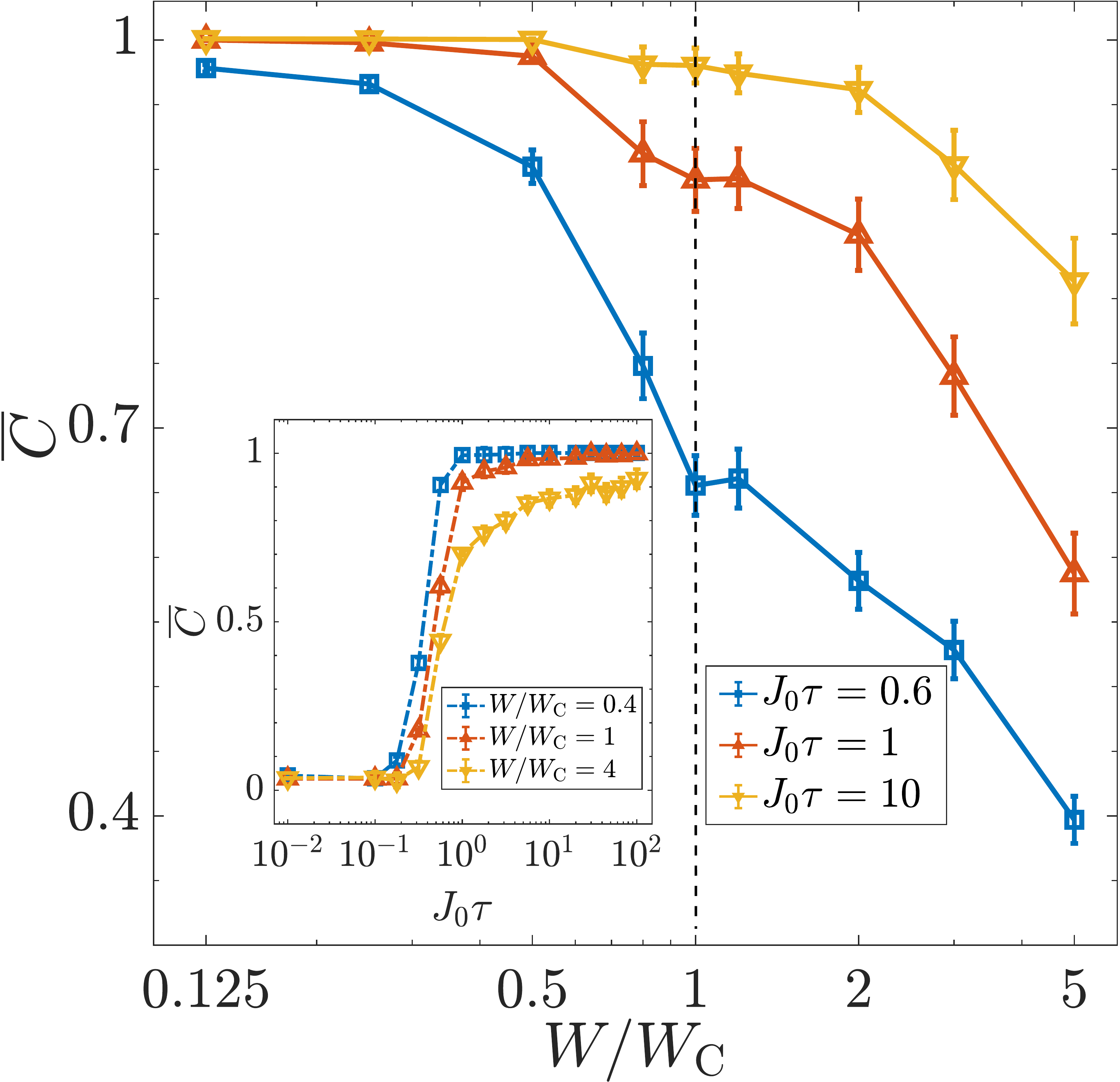}
\caption{The quantum reservoir learning performance on the parity check. 
The learning performance characterized by the normalized covariance $\overline{C}$ monotonically decreases with the disorder strength ($W$), as opposite to the STM task.  The MBL to quantum ergodic phase transition boundary is indicated by the vertical black ``dashed" line. The inset shows the  dependence of $\overline{C}$ on the time duration $\tau$, having a monotonic increasing and saturation behavior.   
In this plot, all parameters are chosen the same as the results of STM task in Fig.~\ref{fig:STM}, except for the time delay 
set by $k_\Delta=4$ here.
}
\label{fig:PC}
\end{figure}

{\it Parity Check Task.---} 
The second learning task we examine with our QRC is PC, for which the learning function is 
${y}^\star _k  = \left(\sum_{m=0}^{k_\Delta}s_{k-m}\right)\,{\rm mod}\, 2$.  This function is highly nonlinear, and the QRC learning performance then reflects the amount of nonlinearity in the reservoir. 
Fig.~\ref{fig:PC} shows our numerical results. As the time duration $\tau$ is increased, we find that $\overline{C}$ monotonically  increases, and then saturates to $1$ at large $\tau$. 
With increasing disorder strength for $J_0 \tau \le 1$, the learning performance becomes slightly worse within the quantum ergodic phase, and bends downward dramatically across the MBL transition---the learning power of the quantum ergodic phase thus outperforms the MBL phase for the PC task. 
In the quantum ergodic phase, the Heisenberg time evolution of the Pauli-$X$ operators is strongly coupled with an exponential number ($4^N-1$) of all Pauli operators, which then resembles the classical chaotic dynamics, providing a larger degree of nonlinearity than the integrable MBL phase.  

\begin{figure}[htp]
\includegraphics[width=0.9\linewidth]{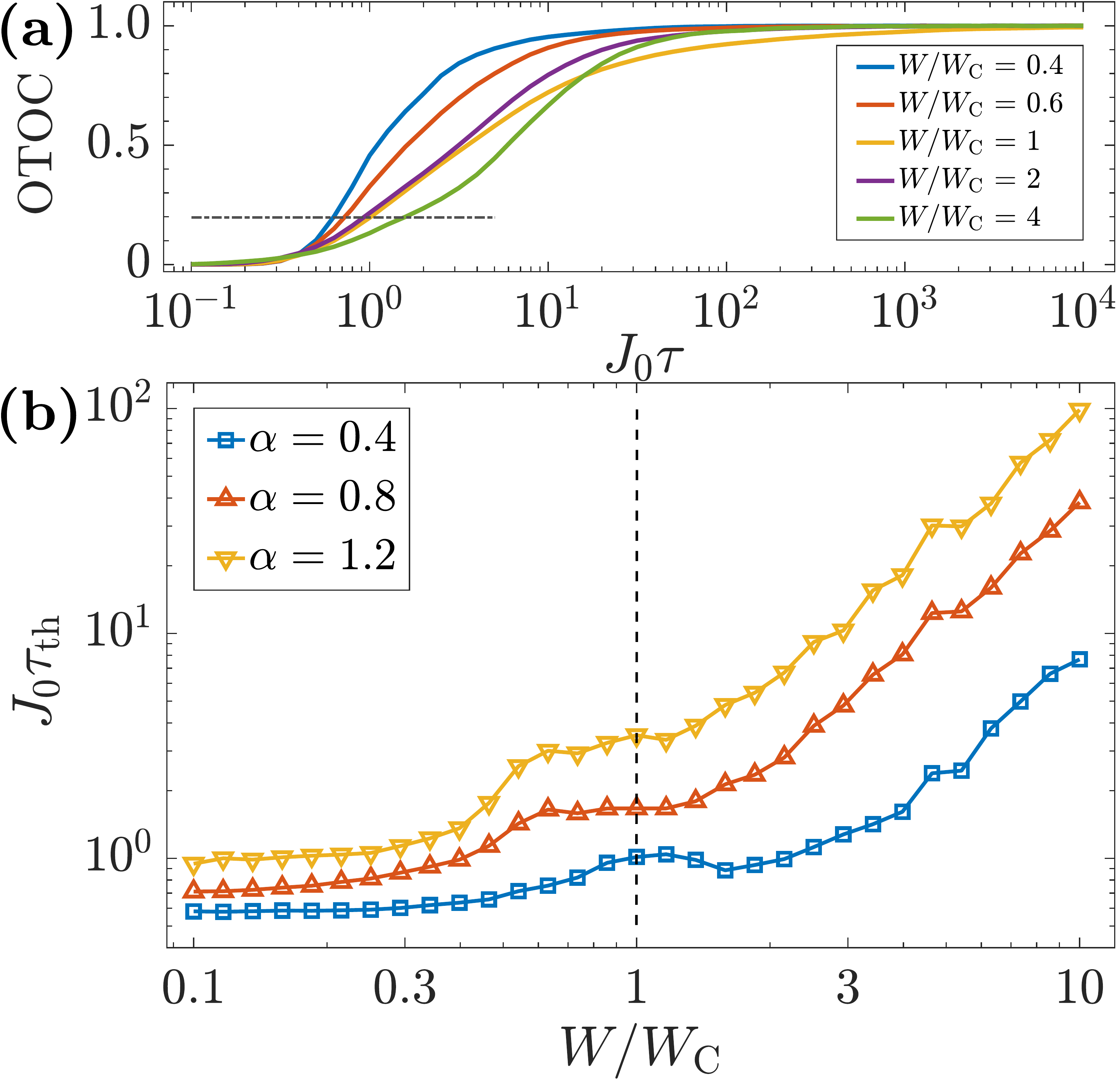}
\caption{Quantum information scrambling with the disordered quantum spin chain reservoir. 
({\bf a}), The averaged OTOC as a function of the time duration  $\tau$ for different disorder strengths ($W$). 
The results are averaged over $100$ random disorder samples. 
In ({\bf a}), the  interaction exponent is fixed at $\alpha=0.4$.  
The horizontal black ``dash-dotted" line marks a threshold OTOC value of $0.2$ (see the main text). A time scale $\tau_{\rm th}$ is defined as the point where OTOC crosses the threshold line. 
({\bf b}), The threshold time $\tau_{\rm th}$ as a function of the disorder strength, with different $\alpha$ exponents.  
The vertical black ``dashed" line indicates the MBL to quantum ergodic phase transition  boundary. 
Here we choose a  qubit number, $N=10$.
}
\label{fig:OTOC}
\end{figure}

To gain more insight about the nonlinearity of the quantum reservoir, we look into the intrinsic chaotic properties of the quantum reservoir and calculate the out-of-time-order-correlator (OTOC)~\cite{1969_Larkin_SPJ,Stanford_JHEP2014}, 
\be
\textstyle O(\tau) = 1- \frac{1}{N-1}\sum_{i=2}^{N}\left\langle X^\dag_1(\tau)X^\dag_iX_1(\tau)X_i\right\rangle\, ,
\ee
with 
$X_1(\tau)=e^{\im H\tau}X_1 e^{-\im H\tau}$ and $\braket{\cdot}$ an average over a thermal ensemble at infinite temperature. 
The OTOC has been introduced in the literature to diagnose quantum chaos by generalizing Poisson brackets and Lyapunov exponent from classical dynamical systems~\cite{2016_Maldacena_JHEP}. 
Here it captures the information spreading from the first qubit to the rest. 
From Fig.~\ref{fig:OTOC} ({\bf a}), we see that OTOC increases monotonically with the time duration $\tau$ and then saturates, which closely correlates with the learning performance in Fig.~\ref{fig:PC}. 
Quantum chaotic dynamics as illustrated by OTOC is beneficial for producing nonlinearity. 
Since all the present NISQ devices have limited quantum coherence time~\cite{Preskill2018quantumcomputingin,2020_Deutsch_PRXQ}, it is crucial to know the required time scale for the QRC  to reach  satisfactory  learning performance. 
We thus introduce a  time scale, $\tau_{\rm th}$, where OTOC exceeds a certain threshold value (set to be $0.2$ here).   
As shown in Fig.~\ref{fig:OTOC} ({\bf b}), we find in general the information spreads much more rapidly in the quantum ergodic phase than the MBL phase. 
We also find the information spreads more rapidly at a smaller $\alpha$, which implies larger amount of nonlinearity  with longer ranged interacting models for a given amount of evolution time. These findings would shed light on quantum reservoir engineering for harnessing the learning power of NISQ devices on nonlinear reservoir computing tasks.

\begin{figure}[htp]
\includegraphics[width=0.9\linewidth]{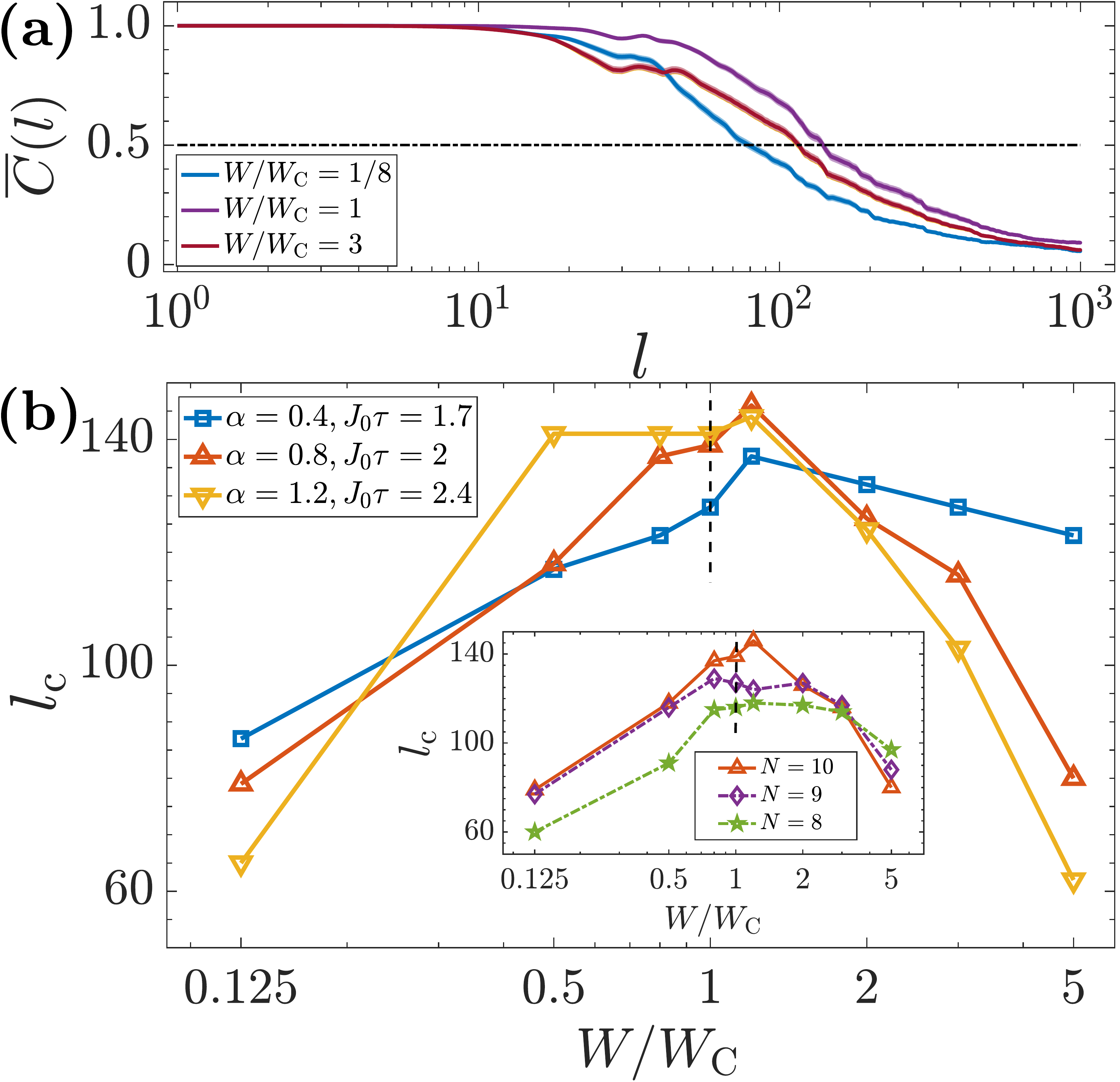}
\caption{The quantum reservoir learning performance on the Mackey glass task. The normalized covariance $\overline{C}$ characterizes the learning performance of this chaotic time sequence prediction task. 
({\bf a}), The dependence of $\overline{C}$ on the number of forward time steps ($l$) for the reservoir computing to predict. Here we set the interaction exponent $\alpha=0.8$, the time duration $J_0\tau=2$, the number of subintervals $V=10$, and the qubit number $N=10$.
The results are calculated by averaging over $300$ samples, with the standard deviations illustrated by the shaded error bands. 
Each learning task instance consists of $1000$ time steps at the beginning discarded, the following $10000$ time steps for training, and the last $l$ time steps for prediction. 
The horizontal black ``dash-dotted" line marks a threshold value of $0.5$ (see the main text). A critical number of forward time steps $l_{\rm c}$ is defined as the point where $\overline{C}$ traverses the threshold value. 
({\bf b}) The critical number $l_{\rm c}$ as a function of disorder strength for different interaction exponents $\alpha$ and time duration $\tau$. 
The inset shows the results with different qubit numbers $N = 8, 9, 10$, where we choose $\alpha=0.8$, and $J_0\tau=2$. 
The vertical black ``dashed" lines in ({\bf b}) indicate  the MBL transition phase boundary. Here, the time durations  $\tau$ for different exponent $\alpha$ are chosen to optimize  the overall critical time step $l_{\rm c}$. 
}
\label{fig:MG}
\end{figure}

{\it Mackey-Glass Task.---} 
With the STM and PC tasks, we have demonstrated maximizing the memory capacity and nonlinearity favors the quantum ergodic, and the MBL phases, respectively.  This suggests the QRC has optimal learning power in between for learning tasks that requires both large memory capacity and sufficient nonlinearity. We thus analyze the MG task~\cite{Jaeger2004, mackey1977oscillation} as a such example. 
This task is defined by~\cite{mackey1977oscillation}, 
\be
\textstyle F_{k+1} = \gamma F_{k}+\frac{\lambda F_{k-k_{\Delta}}}{1+F^{\beta}_{k-k_{\Delta}}},
\ee
with $k_{\Delta}$ a time delay. The dynamical system has a chaotic attractor with $k_{\Delta}>16.8$, and  here we choose $\gamma=0.9$, $\beta=10$, $\lambda=0.2$, and $k_{\Delta}=17$. 
We set $s_k = (F_{k}-F_{\rm min}) /(F_{\rm max} - F_{\rm min}) $, and $y^\star_k = s_{k+1}$---the time sequences are shifted and rescaled  to fit into the regime of $[0, 1]$ by introducing $F_{\rm max}$ and $F_{\rm min}$ in our numerical simulation. 
This learning task is to predict the chaotic time sequence, which is deterministic unlike the two tasks studied above. 
For the MG task, the output signals as measured from the reservoir (Fig.~\ref{fig:QRC}) are added with a slight amount of white noise in the range of $(-\sigma,\sigma)$  to improve robustness~\cite{Keisuke2017}. We choose $\sigma=10^{-6}$ for the noise strength. In predicting the chaotic time sequence, i.e., for $k\in K_3$, $s_k$ and $y_k$ are decided through an iteration---$y_k$ is obtained by giving the input $s_k$ to the reservoir, and then $y_{k+1}$ obtained by setting $s_{k+1} = y_k$. 

The QRC performance on this MG task strongly depends on the number of forward steps ({$l = |K_3|$}) to predict. The dependence of $\overline{C}$ on $l$ is shown in Fig.~\ref{fig:MG} ({\bf a}). The QRC prediction for the chaotic time sequence  works perfectly if the number of forward steps  is small, producing $\overline{C}\approx 1 $ for $l<10$. 
It fails if $l$ at request is too large.  
We then define a critical time step $l_{\rm c}$, above which $\overline{C}$ drops down below a threshold (chosen to be  $0.5$ here). This critical time step quantifies the learning power of the quantum reservoir in predicting the chaotic time sequence. 
As shown in Fig.~\ref{fig:MG} ({\bf b}), as we vary the disorder strength, $l_{\rm c}$ develops a peak generically around the MBL localization transition point, as demonstrated for a broad range of $\alpha\in\left \{0.4, 0.8, 1.2\right \}$. This feature  becomes more prominent for larger number of qubits. This finding implies the quantum reservoir has an optimal learning power near the edge of quantum ergodicity for complex learning tasks that require both memory and nonlinearity being sufficient.

{\it Conclusion and Outlook.---}
We have investigated the learning power of a disordered quantum spin chain in both quantum ergodic and MBL phases. The MBL quantum reservoir is advantageous in holding long-term memory, for the presence of emergent local integrals of motion. The quantum ergodic phase provides a larger degree of nonlinearity, with a compromise on the memory holding. In dealing with the MG task that requires both memory capacity and nonlinearity, we find the learning performance develops a peak near the edge of the quantum ergodic phase, as analogous to the optimal computational power established for the classical reservoir computing  at the edge of chaos. This leads to an important guiding pricinple of quantum reservoir engineering for complex reservoir computing tasks, that is to prepare the quantum system at the edge of quantum ergodicity.

{\it Acknowledgement.---}
This work is supported by National Natural Science Foundation of China (Grant No. 11774067 and 11934002), National Program on Key Basic Research Project of China (Grant No. 2017YFA0304204), and Shanghai Municipal Science and Technology Major Project (Grant No. 2019SHZDZX01), Shanghai Science Foundation (Grant No. 19ZR1471500), the Open Project of Shenzhen Institute of Quantum Science and Engineering (Grant No.SIQSE202002). X.Q. acknowledges support from National Postdoctoral Program for Innovative Talents of China under Grant No. BX20190083.

\bibliography{references}

\end{document}